\documentclass{moriond}

\def\be{\begin{equation}}
\def\ee{\end{equation}}
\def\bea{\begin{eqnarray}}
\def\eea{\end{eqnarray}}

\newcommand{\Photo}{}

\begin{document}
\vspace*{4cm}
\title{Supersymmetry searches at ATLAS and CMS}

\author{ H. Kwon on behalf of the ATLAS and CMS Collaborations }

\address{Department of Physics, University of Antwerp,\\
Prinsstraat 13, 2000 Antwerp, Belgium}

\maketitle

\renewcommand{\thefootnote}{\fnsymbol{footnote}}
\footnotetext[1]{\textcopyright{} Copyright 2025 CERN for the benefit of the ATLAS and CMS Collaboration. CC-BY-4.0 license.}
\renewcommand{\thefootnote}{\arabic{footnote}}

\abstracts{
This document presents recent results from the ATLAS and CMS experiments at the CERN Large Hadron Collider (LHC) on supersymmetry (SUSY) searches. These searches provide results with some non-conventional SUSY search assumptions such as with compressed mass scenario or with long-lived particle scenario. Some of the recent results covers the unexplored SUSY phase space from prior results, brings a new insight to the SUSY landscape. Sensitivity on SUSY upper limits are enhanced with newly added LHC data, enhanced trigger algorithm or advanced event selection with machine learning (ML).
}

\section{Introduction}

At the beginning of the LHC era, the search for SUSY was one of the most important physics programs besides the discovery of the Higgs boson. After years of data taking, there has been no strong evidence in the observation of SUSY particles, with the typical and easiest assumptions. This makes recent SUSY searches refine their assumptions and explore less common phase spaces. 

The prior results on SUSY from LHC experiments are mostly from simplified model assumptions~\cite {LHCNewPhysicsWorkingGroup:2011mji}, which reduces the large number of free parameters in the full model SUSY into a handful of manageable parameters. This enables the experiments to provide primary insights on SUSY sensitivity and make interpretation easy, however more realistic scenario can be considered to provide more precise sensitivity and to find the missing corners. Both ATLAS and CMS Collaborations provided phenomenological MSSM (pMSSM) results fairly recently in the spirit of this~\cite{ATLAS:2024qmx,CMS-PAS-SUS-24-004}.

From a physics signature point of view, both collaborations are providing results with less common signatures, thus technically more challenging. Soft object SUSY searches have been conducted where soft final state objects are expected from the compressed mass spectrum SUSY scenario. Also, long-lived SUSY search results are provided, which complement the typical assumption that SUSY particles decay promptly in the detector. 

Furthermore, results leveraging less common signatures from R-parity violation assumption, flavor-violating interactions~\cite{ATLAS:2024lpr}, and boosted final state objects originating from heavy resonance were reported recently.

\section{Search for electroweak production of supersymmetric particles in scenarios with compressed mass spectra}

A compressed mass spectrum refers to the case where the mass difference of the heavier SUSY particle and the lightest SUSY particle (LSP) is smaller than a few tens of GeV. The target final state objects have very low transverse momentum in this scenario.

The CMS Collaboration recently released two complementary results targeting compressed higgsino scenarios, each focusing on different regions of the mass splitting ($\Delta m(\tilde{\chi}^{\pm}_1 - \tilde{\chi}^0_1)$). 

The first analysis~\cite{CMS-PAS-SUS-24-012} is dedicated to the regime where mass splitting is below 1 GeV. It employs isolated soft tracks using a parametrized neural network (PNN) for track tagging. The result achieves a sensitivity to soft, disappearing track signatures. This analysis excludes higgsino masses up to 195 GeV in this highly compressed region. 

The second analysis~\cite{CMS-PAS-EXO-23-017} extends the sensitivity to sub-GeV mass splittings, reaching down to 0.5 GeV. Leveraging improved object reconstruction through machine learning, it enables the identification of electrons with transverse momenta as low as 1 GeV, enabling access to previously unexplored phase space. 

\begin{figure}
    \centering
    \includegraphics[width=0.32\textwidth]{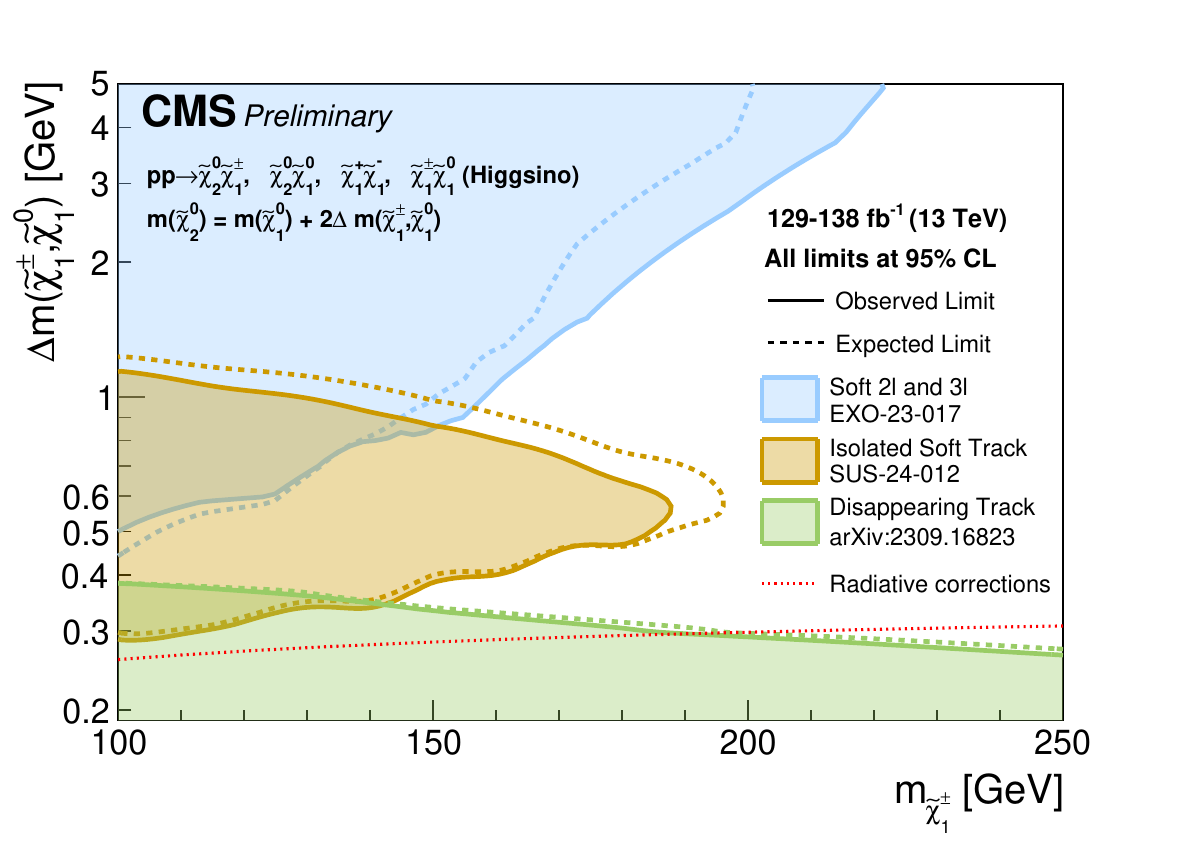}
    \includegraphics[width=0.32\textwidth]{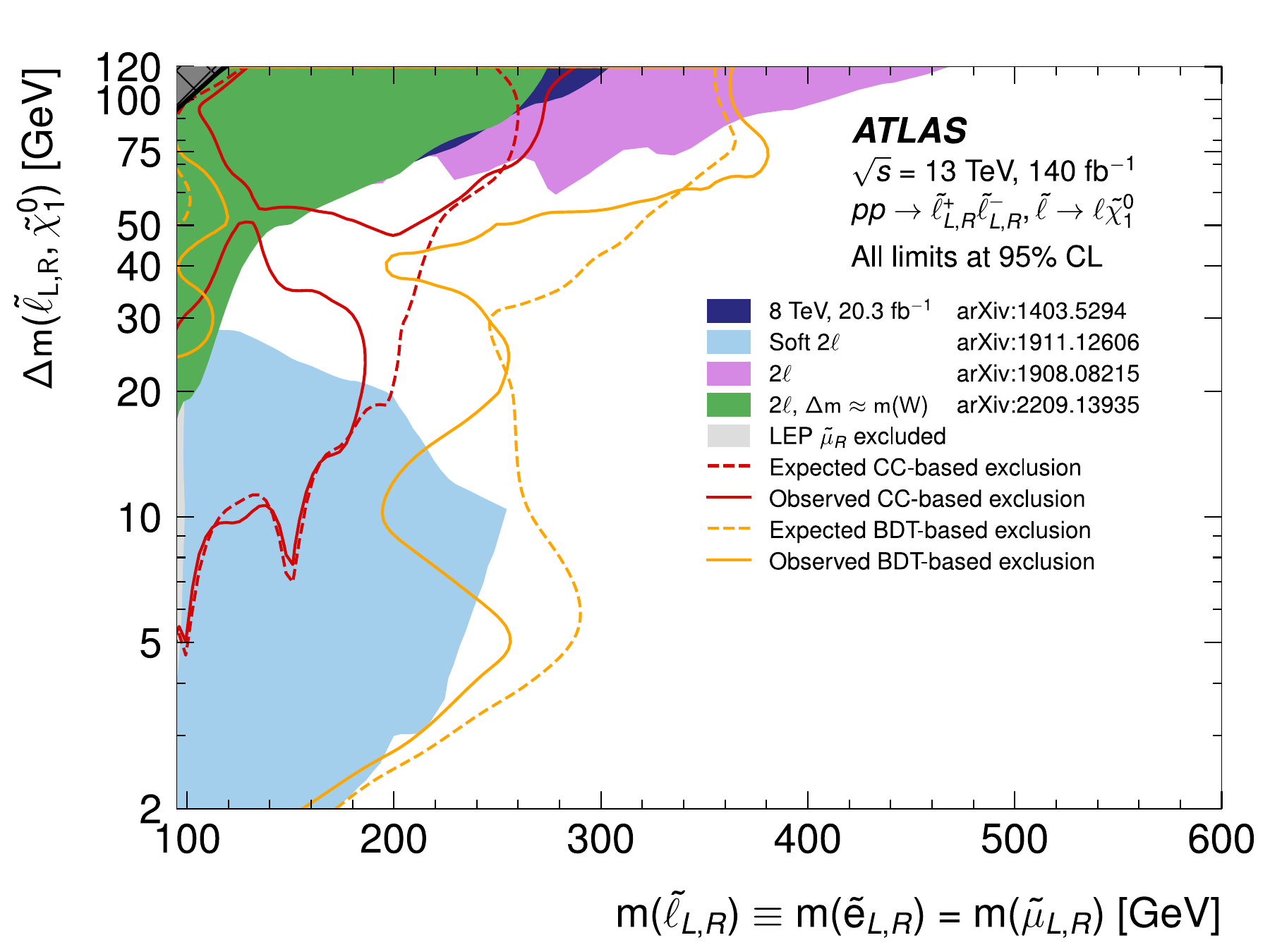}
    \includegraphics[width=0.32\textwidth]{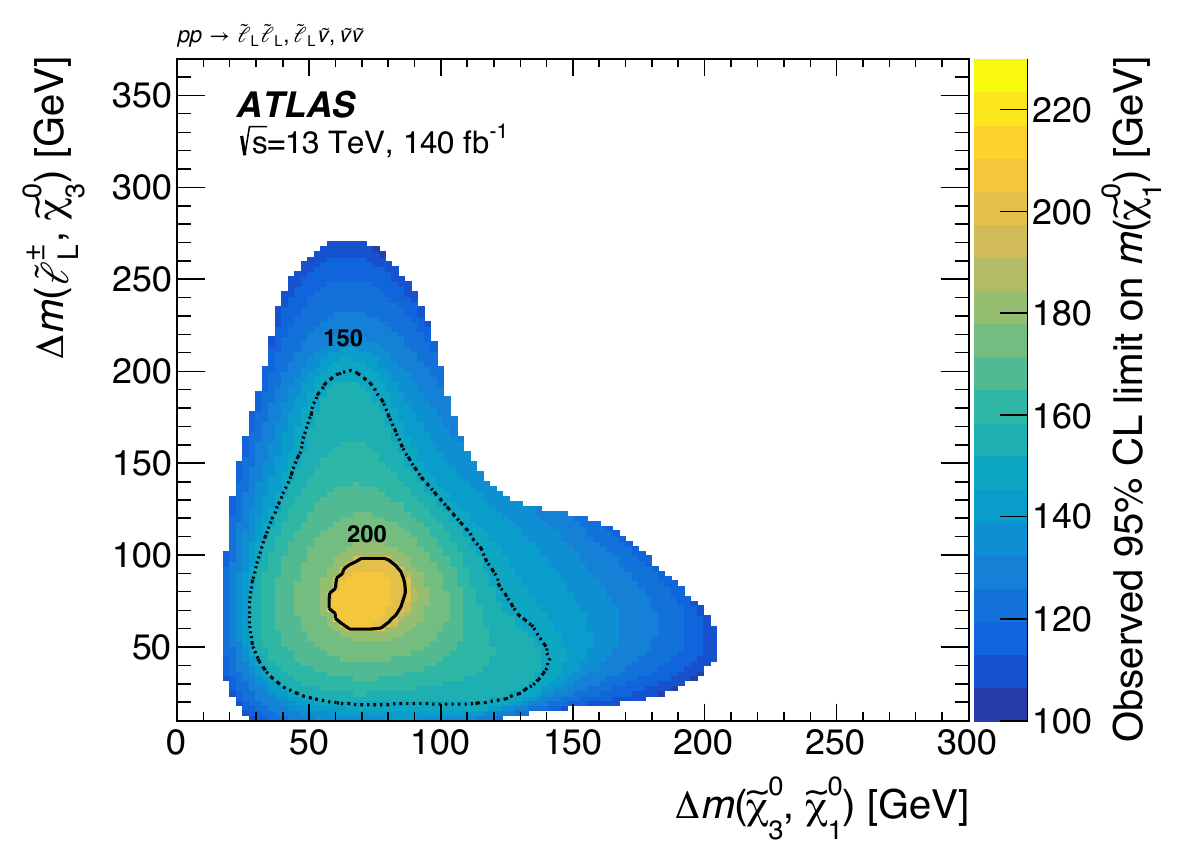}
    \caption{(Left) Exclusion contours for higgsino in $\Delta m(\tilde{\chi}^{\pm}_1 - \tilde{\chi}^0_1)$, higgsino mass plane from three analyses. The upper two sets of contours were added recently. (Middle) Direct slepton production search. Contours from this analysis are in yellow and red lines. Color-shaded contours are from prior analyses. (Right) Slepton cascade decay search. Observed upper limits are shown in $\Delta m(\tilde{l}^{\pm} - \tilde{\chi}^0_3)$, $\Delta m(\tilde{\chi}^0_3 - \tilde{\chi}^0_1)$ plane.}
    \label{fig:figure1}
\end{figure}

Together, these analyses provide complementary coverage of the compressed higgsino parameter space. As shown in Fig.~\ref{fig:figure1} (left), CMS now probes the full range of $\Delta m(\tilde{\chi}^{\pm}_1 - \tilde{\chi}^0_1)$.

The ATLAS Collaboration released a result on direct slepton production in a compressed mass spectrum scenario~\cite{ATLAS:2025evx}. The analysis utilizes an ML approach to improve sensitivity. Figure~\ref{fig:figure1} (middle) presents the exclusion limits for a four-fold degenerate slepton scenario (selectrons and smuons, both left and right-handed), showing the most stringent limits to date in this phase space. A mild excess observed in the data leads to a weaker observed exclusion compared to the expected limit in the region around 40 GeV. Despite this, slepton masses are excluded up to approximately 400 GeV.

Another ATLAS result on direct slepton production sets mass limits within a specific SUSY model that includes an intermediate bino in the decay chain~\cite{ATLAS:2025dns}. Figure~\ref{fig:figure1} (right) shows the analysis sensitivity as a function of both the mass difference between the slepton and the bino, and between the bino and the LSP. The analysis provides unique sensitivity to the region where $\Delta m(\tilde{\chi}^0_3 - \tilde{\chi}^0_1) > 100$ GeV.

\section{Search for long-lived supersymmetric particles}

As one possible scenario in which SUSY has remained undetected at the LHC, SUSY particles with a long-lived nature which may escape detection have been investigated.

The ATLAS Collaboration released a displaced lepton analysis targeting long-lived sleptons and charginos~\cite{ATLAS:2024vnc}. The analysis makes use of newly collected Run 3 data, a dedicated trigger optimized for large impact parameters and low transverse momentum leptons. It also employs complementary signal regions in the electron channel using a boosted decision tree to enhance sensitivity. Figure~\ref{fig:figure2} (left) shows that this analysis improves the previous result, due to both increased luminosity and the enhanced sensitivity via a dedicated trigger algorithm.

\section{Search for electroweakino production mediated by a heavy resonance}

Typical direct SUSY pair production searches target final states of objects with modest to medium transverse momentum. However, if the SUSY production is mediated by a heavy resonance, the final objects are boosted and have much larger transverse momentum, which narrows down the phase space of interest to much boosted regions. A SUSY extension with a larger theoretical framework expects such a new particle alongside SUSY particles.

The CMS Collaboration has recently released the first direct search targeting such an extended model, where supersymmetric particles are produced via a heavy neutral gauge boson Z'~\cite{CMS-PAS-SUS-23-006}. This analysis exploits the unique topology of the signal via PNN. The result provides novel interpretations: it sets limits on the chargino mass produced via a heavy resonance, and it constrains the mass of the Z' boson via the SUSY decay channel, offering a complementary probe to conventional resonance searches. Figure~\ref{fig:figure2} (middle) shows upper limits from this analysis.

\section{Search for R-parity violating supersymmetric signatures}

With R-parity-violating (RPV) scenarios, the target signature can be unique compared to the conventional R-parity-conserving SUSY searches, such as little or no missing transverse momentum due to the scenario that the LSP is unstable within the assumption.

\begin{figure}
    \centering
    \includegraphics[width=0.32\textwidth]{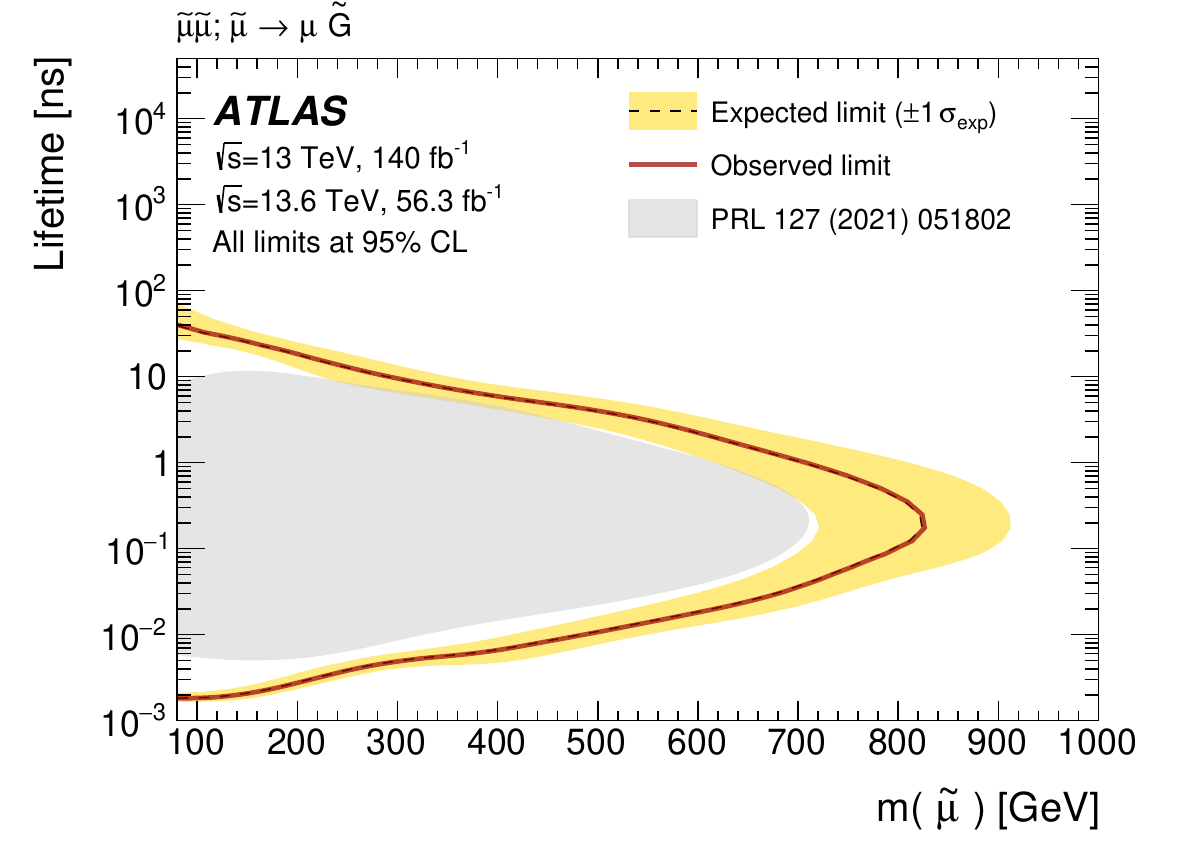}
    \includegraphics[width=0.32\textwidth]{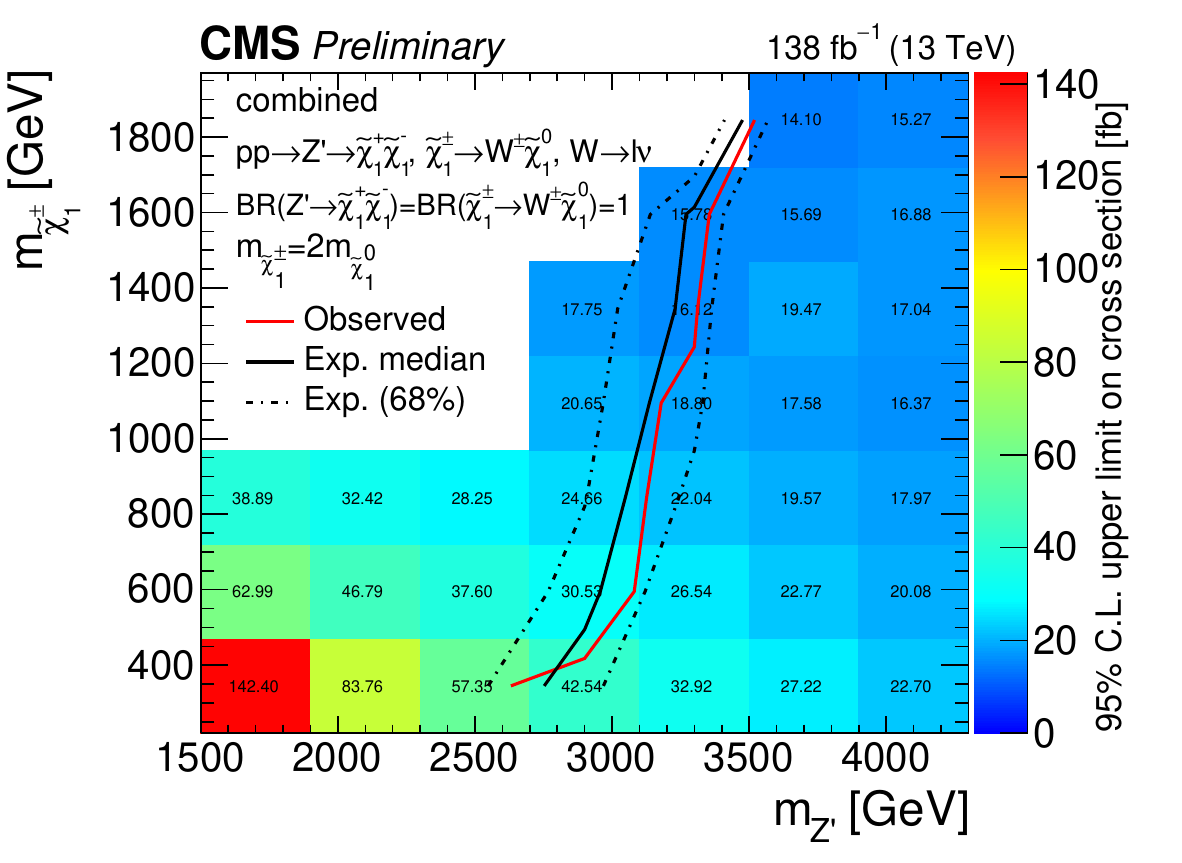}
    \includegraphics[width=0.32\textwidth]{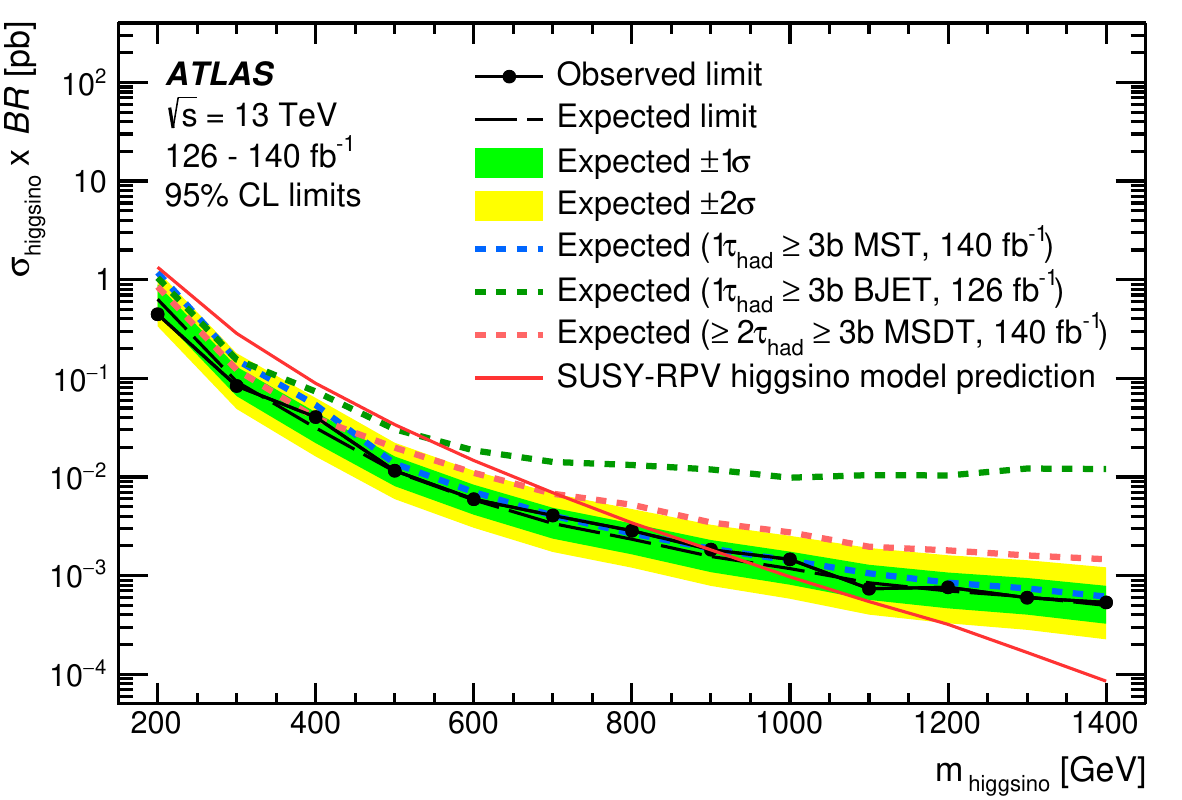}
    \caption{(Left) Displaced leptons search. Exclusion contours for smuon and its lifetime. The gray shaded area is from a prior analysis. (Middle) Search for Z' in charginos decay. Upper limits on cross section in Z', chargino mass plane. (Right) Search for electroweak production in third-generation final states. Upper limits with RPV higgsino model interpretation.}
    \label{fig:figure2}
\end{figure}

The ATLAS Collaboration has recently released an analysis interpreted in the context of RPV SUSY, targeting final states involving third-generation particles~\cite{ATLAS:2025wgc}. The signature includes multiple hadronically decaying taus and jets originating from b-quarks. The analysis employs a PNN to enhance signal discrimination. In the interpretation within higgsino and wino simplified models, the analysis sets exclusion limits exceeding 800 GeV for the masses of both the higgsino and wino. Figure~\ref{fig:figure2} (right) shows constrains on higgsino mass with RPV interpretation.

\section{Summary }

Recent results featuring rather non-conventional SUSY assumptions and novel methodologies are presented. Search for SUSY remains an active part of physics programs in ATLAS and CMS Collaborations, and those results unveiled new SUSY landscape. Further possible corners of SUSY phase spaces are to be studied.

\section*{Acknowledgments}

H. Kwon would like to thank the ATLAS and CMS Collaborations and the organizers of the 59th Recontres
de Moriond for the opportunity to present these results.



\section*{References}
\bibliography{moriond}


\end{document}